\documentclass[slac_one]{revtex4}
\usepackage{graphicx}
\usepackage{fancyhdr}

\usepackage{bm}
\usepackage{epsfig}
\usepackage{graphics}

\def\itbf#1{\mbox{\boldmath $#1$}}

\def\Dsl{\hbox{/\kern-.6000em D}} 

\def\dsl{\,\raise.15ex\hbox{/}\mkern-13.5mu D}

\def\psip#1{\psi_{\mathbf{#1}}}
\def\chip#1{\chi_{\mathbf{#1}}}

\def\ltap{\ \raise.3ex\hbox{$<$\kern-.75em\lower1ex\hbox{$\sim$}}\ }
\def\gtap{\ \raise.3ex\hbox{$>$\kern-.75em\lower1ex\hbox{$\sim$}}\ }
\def\OMIT#1{}

\def\lsim{\mathrel{\raise.3ex\hbox{$<$\kern-.75em\lower1ex\hbox{$\sim$}}}}
\def\gsim{\mathrel{\raise.3ex\hbox{$>$\kern-.75em\lower1ex\hbox{$\sim$}}}}

\def\msb{{\overline{\rm MS}}}

\newcommand{\bmk}{\mathbf k}
\newcommand{\bmp}{\mathbf p}
\newcommand{\bbmp}{\mbox{\scriptsize\boldmath $p$}}

\newcommand{\bmA}{\mathbf A}

\newcommand{\bmD}{\mathbf D}

\catcode`\@=11
\def\slash{\mathpalette\make@slash}
\def\make@slash#1#2{\setbox\z@\hbox{$#1#2$}%
  \hbox to 0pt{\hss$#1/$\hss\kern-\wd0}\box0}
\catcode`\@=12 

\begin{document}

\title{{\small{2005 International Linear Collider Workshop - Stanford,
U.S.A.}}\\ 
\vspace{12pt}
Effects of Finite Top Lifetime at the $t \bar t$ Threshold} 

%

\author{Andr\'e~H.~Hoang and Christoph~Rei\ss er}
\affiliation{Max-Planck-Institute for Physics, Munich, Germany}

\begin{abstract}
In this talk electroweak corrections related to the finite top quark
lifetime to the total top pair threshold $e^+e^-$ cross section at
NNLL order are discussed. We include the absorptive parts in electroweak
matching conditions of the NRQCD operators related to the top decay
and use the optical theorem. Gauge invariance is maintained. The
corrections lead to ultraviolet phase space divergences and NLL mixing
effects. The corrections can amount to several percent
and are phenomenologically relevant. 
\end{abstract}

\maketitle

\thispagestyle{fancy}


\section{Introduction}
\label{sectionintroduction}
Because in the Standard Model the top quark width
$\Gamma_t\approx 1.5$~GeV is much larger than the typical
hadronization energy $\Lambda_{\rm QCD}$, the total cross section
$\sigma(e^+e^-\to t\bar t)$ in the threshold region $\sqrt{s}\approx
2m_t$ can be computed with perturbative methods to very high
precision. The rapid rise of the cross section in the threshold region
will allow for a measurement of the top quark mass (in a threshold
mass scheme~\cite{synopsis}) with experimental and theoretical
uncertainties at the level of only $100$~MeV. Other parameters such as
$\alpha_s$, $\Gamma_t$ or (if the Higgs boson is light) the top Yukawa
coupling  $g_{\rm tth}$ can also be determined if the normalization
and the exact form of the line-shape can be computed with small
theoretical errors at the percent level~\cite{TTbarsim,synopsis}.
This is required because the observable cross section is a convolution
of the theory prediction with the partly machine-dependent $e^+e^-$
luminosity spectrum~\cite{TTbarsim}.

The theoretical instrument to make first principles predictions for
top threshold observables is based on NRQCD, an effective theory (EFT) of
QCD for heavy quark pairs with small relative velocity
$v\sim\alpha_s\ll 1$. Within ``velocity'' NRQCD
(vNRQCD)~\cite{LMR,HoangStewartultra}, which we will use here, it is
possible to sum Coulomb singularities and (using renormalization group
methods) logarithms of $v$ to all orders of QCD perturbation theory
while a systematic and coherent $v$ power counting is maintained. The 
summation of logarithms avoids large normalization
uncertainties~\cite{hmst,hmst1} that were obtained in earlier
fixed-order predictions~\cite{synopsis}. At NNLL order (i.e. including
corrections of order $v^2$ and the summation of terms
$\alpha_s^2(\alpha_s\ln v)^n$) all QCD ingredients for the threshold
cross section are presently known except for still missing  
subleading mixing effects in the running of the heavy quark pair production
current. The current normalization QCD uncertainties of the total cross
section are estimated to be around $6\%$~\cite{HoangEpiphany}. 

In this talk we are interested in electroweak effects. At leading
order, the three basic electroweak effects are the $t\bar t$
production process itself, the finite top lifetime and the luminosity
spectrum mentioned above. The latter is accounted for in the experimental
simulations~\cite{TTbarsim}, the two former effects are described
by NRQCD. Here we want to discuss finite top lifetime effects at the
subleading level including QCD interference effects. Previous partial
analyses have indicated that the corrections can reach the level of a
few percent~\cite{GuthKuehn,HoangTeubnerdist}. In particular, we 
investigate the role of absorptive parts related to 
the top quark decay in electroweak loop corrections to the NRQCD
matching conditions that contribute to the NNLL total cross section.   
Interestingly, these corrections lead to UV phase space divergences
that would not exist for stable top quarks and that lead to
a NLL renormalization of $(e^+e^-)(e^+e^-)$ operators that also
contributes to the total cross section. For details of our analysis we
refer the interested reader to Ref.~\cite{HoangReisser}.

\section{Power Counting and Matching Conditions} 
\label{sectionpowercounting}

Let us first recall the power counting to classify the order at which 
electroweak effects can contribute by considering the matching conditions
for the vNRQCD bilinear quark field operators. Electroweak corrections
are obtained by matching 2-point functions in the
effective theory to those in QCD {\it and} the electroweak theory. The
NNLL result including also all QCD contributions has the
form~\cite{HoangTeubnerdist}  
\begin{eqnarray} \label{Lke}
 {\mathcal L}_{\rm bilinear}(x) &=& \sum_{\bbmp}
   \psip{\bbmp}^\dagger(x)   \biggl\{ i D^0 - {(\itbf{p}-i\bmD)^2 \over 2 m_t} 
   +\frac{{\itbf{p}}^4}{8m_t^3}  
   + \frac{i}{2} \Gamma_t \bigg( 1 - \frac{{\itbf{p}}^2}{2 m_t^2} \bigg) 
   - \delta m_t \biggr\} \psip{\bbmp}(x) + (\psip{\bbmp} \to\chip{\bbmp})\,,
\end{eqnarray}
where the fields $\psip{\bbmp}$ and $\chip{\bbmp}$ destroy top and
antitop quarks with momentum ${\itbf{p}}$ and {\it positive} energies, 
$D^\mu=(D^0,-\bmD)=\partial^\mu + i g A^\mu$ is the ultrasoft gauge
covariant derivative and $\Gamma_t$ is the top quark width defined at
the (electroweak) top quark pole. The $v$-counting is $D^0\sim m_t
v^2\sim\Gamma_t\sim m_t g^2$ since the top width is of order of the
typical top kinetic energy, which defines the ultrasoft scale $m_t
v^2$. This leads to the counting  
\fbox{$v\,\sim\,\alpha_s\,\sim\,g\,\sim\,g^\prime$}
for the SU(2) and U(1) gauge couplings $g$ and $g^\prime$, where we
can treat the weak mixing being of order one.
The term $\delta m_t$ is a residual mass term of order $v^2$ that arises within
threshold mass schemes~\cite{synopsis}. Its electroweak contributions
are straightforward to compute and will not be discussed here
further. The term $i\Gamma_t(1-\bmp^2/2m_t^2))$ describes the finite
lifetime at LL order and the NNLL time dilatation effects. 
It produces the known replacement rule $E\to
E+i\Gamma_t$~\cite{Fadin1} to account  
for finite lifetime effects at LL order. Although it renders
the effective Lagrangian formally non-hermitian (since hermitian
conjugation does {\it not} change its sign), the EFT inherits
unitarity from the underlying theory, so we can later use the optical
theorem to compute the total rates. It is 
crucial to understand that the effects of the top decay represent hard
physics that can be integrated out (i.e. treated by the matching
condition $i\Gamma_t$) and that, therefore, one can determine the
corresponding EFT matching conditions for on-shell top quark
amplitudes. This is analogous to the treatment of photons in
an absorptive medium in the optical theory. 
So to account for all NNLL order electroweak effects to 
Eq.~(\ref{Lke}) one still has to include the one-loop electroweak
and the ${\cal O}(\alpha_s)$ and ${\cal O}(\alpha_s^2)$ QCD 
corrections to the on-shell top width $\Gamma_t$ which will also not
be discussed here any further. 

Concerning the $t\bar t$ interaction at LL order we only
need to consider the Coulomb potential, 
$
{\cal L}_{\rm pot} = 
\sum_{\bbmp,\bbmp^\prime} \frac{4\pi C_F \alpha_s(m_t\nu)}{(\bmp-\bmp^\prime)^2}\,
\psip{\bbmp^\prime}^\dagger \psip{\bbmp} \chip{-\bbmp^\prime}^\dagger\chip{-\bbmp}
$,
where $\nu$ is the vNRQCD renormalization scaling parameter.
Since effects from the top decay are hard ($\sim m_t$), we can neglect
the momentum exchange $\bmp-\bmp^\prime$ in the electroweak loops
during the matching computation. From this it is easy to see that all the
$g^2$ (vertex and wave function) corrections to the Coulomb
potential cancel due to SU(3) gauge invariance. Since all
electroweak corrections to the $1/m_t$ suppressed potentials are
beyond NNLL order simply by counting powers of $v$ and $g$, we also 
don't have to consider any electroweak corrections to these
potentials. 
The same argument applies to electroweak corrections to
the interactions of top quarks with the soft gluons ($k\sim m_t v$) or
potential four-quark operators caused by electroweak box diagrams; 
the former can only contribute to ${\cal O}(\alpha_s)$ corrections to
the potentials, and the latter are already $1/m_t^2$ suppressed without
even accounting for the powers of $g$. Moreover, the Coulomb interaction
does not generate any UV divergences, so electroweak NNLL
contributions due to mixing do also not exist. 

The dominant operators used to describe
$t\bar t$ spin-triplet production have the form  
\begin{eqnarray}
{\cal O}_{V,\bbmp}  =  
\big[\,\bar e\,\gamma_j\,e\,\big]\,{\cal O}_{\bbmp,1}^j
\,,\qquad
{\cal O}_{A,\bbmp}  = 
\big[\,\bar e\,\gamma_j\,\gamma_5\,e\,\big]\,{\cal O}_{\bbmp,1}^j
\,,
\end{eqnarray}
where
${\cal O}_{\bbmp,1}^j  = 
\Big[\,\psi_{\bbmp}^\dagger\, \sigma_j (i\sigma_2)
  \chi_{-\bbmp}^*\,\Big]$.
They give the contribution 
$
\Delta {\cal L} = \sum_{\bbmp} \left(C_V {\cal O}_{V,\bbmp} +  C_A
{\cal O}_{A,\bbmp} \right) + \mbox{h.c.}
$
to the effective theory Lagrangian where the hermitian conjugation
(which gives the corresponding  annihilation operators) is referring to the
operators, but does not affect their Wilson coefficients. 
Since we neglect QED radiative corrections, the $e^\pm$ fields act
like classic fields, but we need them due to electroweak gauge
invariance. The leading order matching conditions to $C_{V/A}$ are
obtained from the full theory Born diagrams with photon and Z exchange
and are of order $g^2$. We will see in
Sec.~\ref{sectionmatchingconditions} that $C_{V/A}$ will receive  
imaginary matching conditions from one-loop electroweak corrections
that contribute at NNLL order and play a role similar to the imaginary
terms in Eq.~(\ref{Lke}).  

It has been shown in Refs.~\cite{Fadin2} that up to NLL order for
$\sigma_{\rm tot}$ there are no QCD interference effects from
(ultrasoft) gluon radiation off the top/antitop quark or its decay
products. (For this consideration only ultrasoft gluons are relevant
because they cannot kick a top quark off-shell.) The proof was
conducted by explicit computation of 
diagrams. Using the analysis of matching conditions we can show that
this statement even holds at NNLL order. For the time-like ultrasoft
$A^0$ gluons, QCD gauge invariance ensures that the dominant
electroweak matching corrections to the $A^0$ interaction vertex shown
in Eq.~(\ref{Lke}) vanish  because we can again approximate the
ultrasoft gluon momentum in the electroweak loop as zero. Moreover, the
exchange of time-like ultrasoft $A^0$ gluons does itself not
contribute at LL because they can be removed from the LL
particle-antiparticle sector in  Eq.~(\ref{Lke})
by a redefinition of the top and antitop fields related
to static Wilson lines~\cite{Korchemsky1}. An anomalous interaction in
analogy to the $g-2$ is not covered by this argument but suppressed
by a factor $1/m_t^2$. Accounting also for powers of $g$ one finds
that interference from $A^0$ gluons does not contribute at NNLL order.
For the space-like ultrasoft $\bmA$ gluons the conclusion is the same
because they couple to the quarks with the $\bmp.\bmA/m_t$ coupling. The  
$g^2$-suppression from an additional electroweak loop correction to the
interaction vertex then leads to a contribution beyond the NNLL order.

\section{Absorptive Matching Conditions} 
\label{sectionmatchingconditions}
%
%

%
%
\begin{figure}[t] 
\includegraphics[width=8cm]{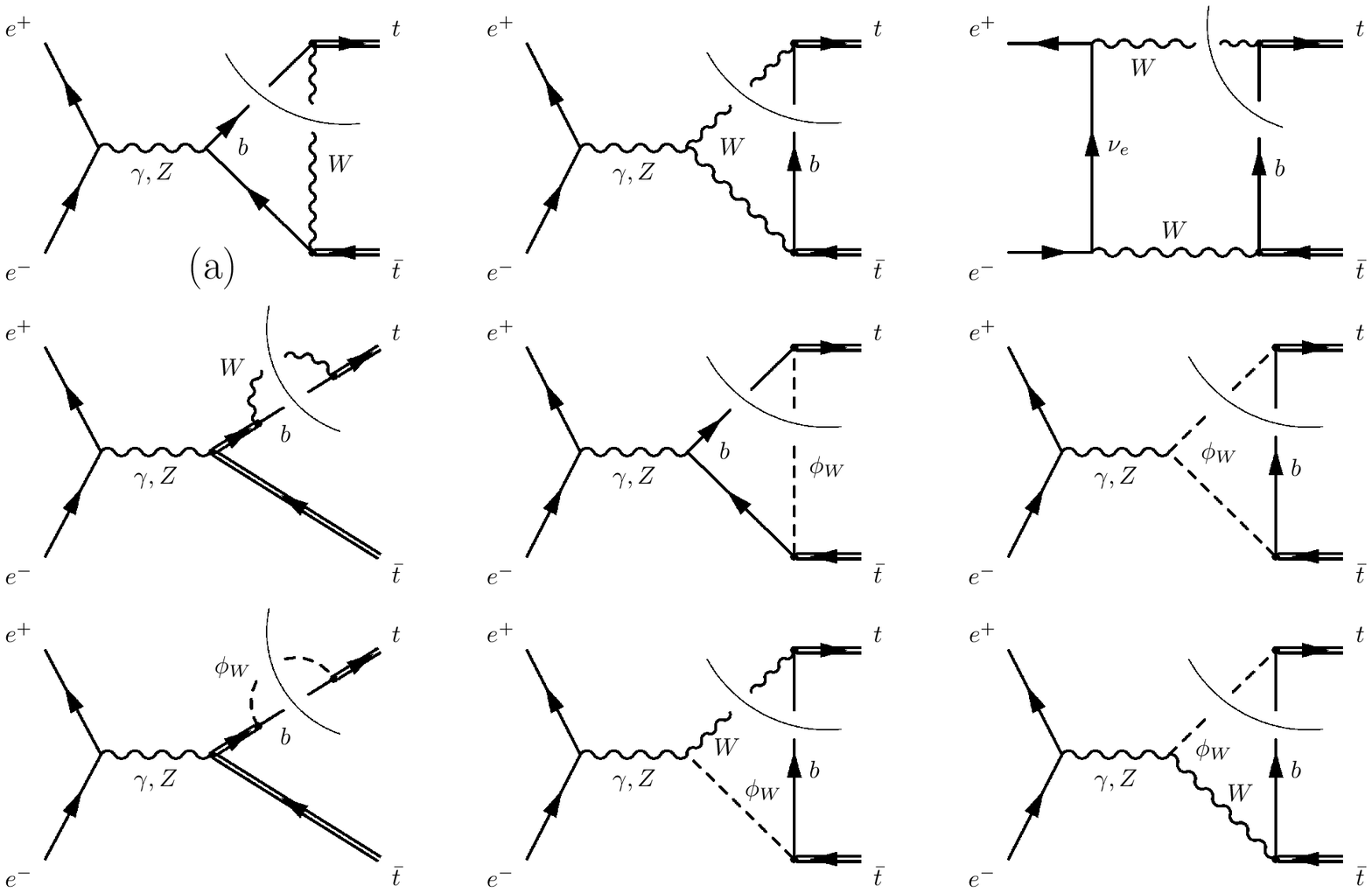}
\qquad
\includegraphics[width=8cm]{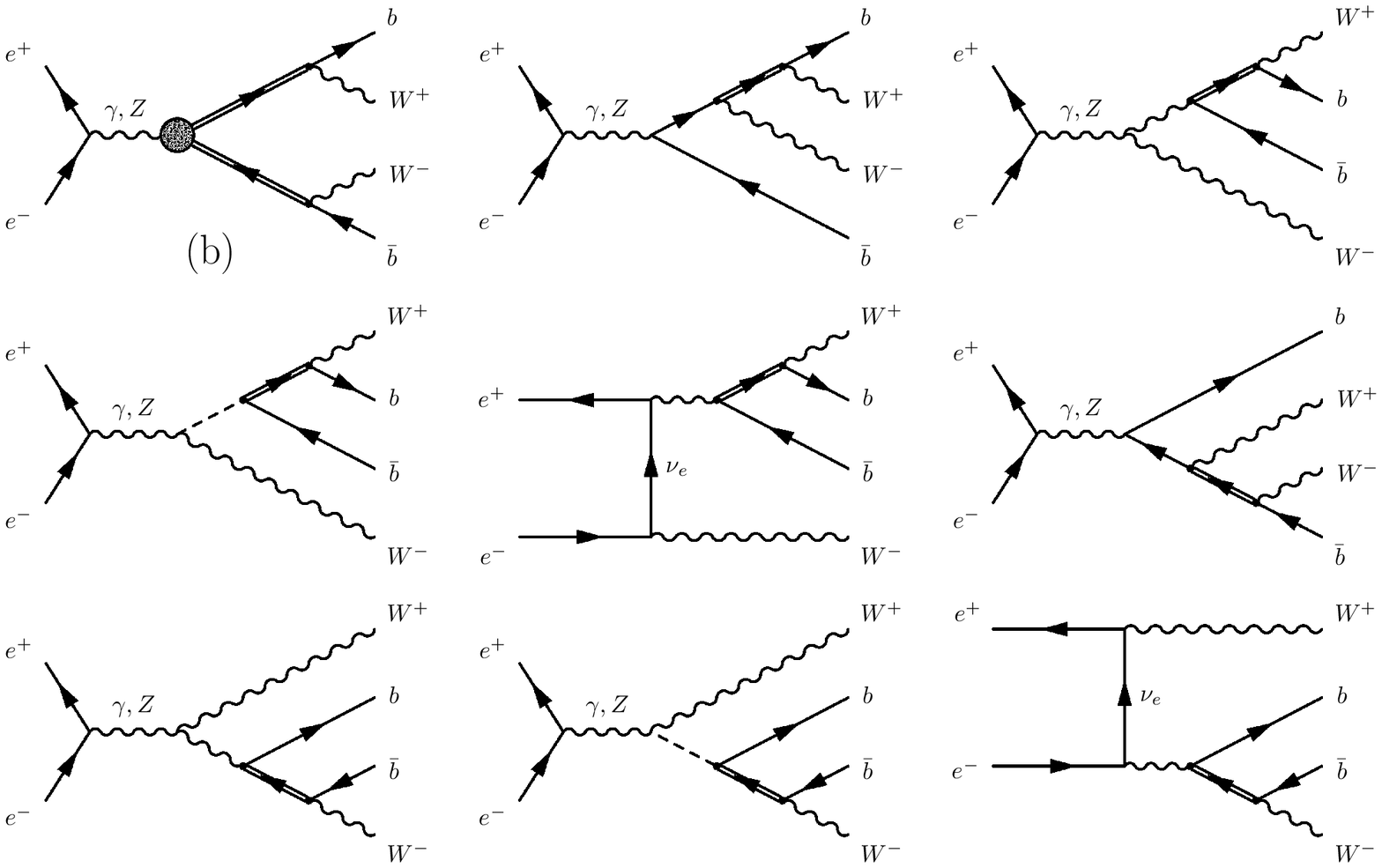}
 \caption{
(a) Full theory diagrams in Feynman gauge needed to determine the 
electroweak absorptive parts in the Wilson coefficients $C_{V/A}$  
related to the physical $bW^+$ and $\bar b W^-$ intermediate states. Only the
$bW^+$ cut is drawn explicitly. 
(b) 
Full theory diagrams describing the process 
$e^+e^-\to bW^+\bar b W^-$ with one or two intermediate top or antitop quark
propagators. The circle in the first diagram represents the QCD form
factors for the $t\bar t$ vector/axial-vector currents.
 \label{fig1} }
\end{figure}

To determine the absorptive parts related
to the top decay of the matching conditions of the
operators ${\cal O}_{V,\bbmp}$ and ${\cal O}_{A,\bbmp}$ that
contribute to the total cross section at NNLL order, we have to
consider the $bW^+$ and $\bar b W^-$ cuts of the full theory diagrams
shown in Fig.\,\ref{fig1}a. To obtain the contributions at this order
the external (on-shell) top quarks can be taken to be at rest. The
full theory amplitude has the form 
\begin{eqnarray}
{\cal A} & = &
i\,\Big[\,\bar v_{e^+}(k^\prime)\,\gamma^\mu
 (i C_V^{\rm bW,abs}+i C_A^{\rm bW,abs}\,\gamma_5)\, u_{e^-}(k)\,\Big]\,
\Big[\,\bar u_t(p)\,\gamma_\mu\,v_{\bar t}(p)\,\Big]
\,,
\label{eett}
\end{eqnarray}
where $k+k^\prime=2p=(2m_t,0)$. The amplitude of the charge conjugated
process describing top pair annihilation reads
\begin{eqnarray}
\bar{\cal A} & = &
i\,\Big[\,\bar u_{e^-}(k)\,\gamma^\mu
 (i C_V^{\rm bW,abs}+i C_A^{\rm bW,abs}\,\gamma_5)\, v_{e^+}(k^\prime)\,\Big]\,
\Big[\,\bar v_{\bar t}(p)\,\gamma_\mu\,u_t(p)\,\Big]
\,.
\label{eettadj}
\end{eqnarray}
We used the cutting equations to obtain expressions for $i
C_{V/A}^{\rm bW,abs}$ 
and checked electroweak gauge invariance by carrying out the computation in
unitary and Feynman gauge. Analytic formulae are provided in
Ref.~\cite{HoangReisser}. The results are consistent with results
obtained earlier in Ref.~\cite{GuthKuehn}.
It is a consequence of the unitarity of the underlying theory that the
sign of the imaginary part of the amplitude does not change in the
charge conjugated amplitude.
It is straightforward to match the amplitudes for the operators ${\cal O}_{V/A,\bbmp}$
and ${\cal O}^\dagger_{V/A,\bbmp}$ to the full theory results in Eqs.\,(\ref{eett})
and (\ref{eettadj}).
The resulting matching conditions for the operators ${\cal O}_{V/A,\bbmp}$
and ${\cal O}^\dagger_{V/A,\bbmp}$  read
\begin{eqnarray}
C_{V/A}(\nu=1) & = & C_{V/A}^{\rm born} + i\,C_{V/A}^{\rm bW,abs}
\,,
\label{Cvaabs}
\end{eqnarray}
where we have also indicated the Born level contributions.
In a full treatment of electroweak effects the coefficients $C_{V/A}$
also include the real parts of the full set of electroweak one-loop
diagrams  indicated in Fig.\,\ref{fig1}a. A comprehensive examination of
these contributions will be provided in subsequent analyses.

\section{Time-Ordered Product and Renormalization} 
\label{sectionrenormalization}

Through the optical theorem the NNLL order corrections to the total cross
section that come from the absorptive one-loop electroweak matching conditions
for the operators ${\cal O}_{V/A,\bbmp}$ and from the time dilatation corrections
can be computed from the 
imaginary part of the $(e^+e^-)(e^+e^-)$ forward scattering amplitude,
\begin{eqnarray}
\sigma_{\rm tot} & \sim &
\frac{1}{s}\, 
\mbox{Im}\left[\,\Big(C_V^2(\nu)+C_A^2(\nu)\Big)\,L^{lk}\,{\cal  A}_1^{lk}\,\right]
\,,
\end{eqnarray}
where $L^{lk} = \frac{1}{2}\,(k+k^\prime)^2\,(\delta^{lk}-\hat e^l
\hat e^k)$ is the spin-averaged lepton tensor and 
\begin{eqnarray}
{\cal A}_1^{lk} = 
i\, \sum\limits_{\mbox{\scriptsize\boldmath $p$},\mbox{\scriptsize\boldmath $p'$}} \int\! d^4x\:
e^{-i\hat q \cdot x}\:
\Big\langle\, 0\,\Big|\,T\,
{{\cal O}_{\mbox{\scriptsize\boldmath $p$},1}^l}^{\!\!\!\dagger} (0)\, 
{\cal O}_{\mbox{\scriptsize\boldmath $p'$},1}^k (x)\Big|\,0\,\Big\rangle
 = 2\,N_c\,\delta^{lk}\,G^0(a,v,m_t,\nu)
\end{eqnarray}
is the time-ordered product of the $t\bar t$ production and
annihilation operators ${\cal O}_{\bbmp,1}^j$ and ${{\cal
    O}_{\bbmp,1}^j}^{\!\!\!\dagger}$~\cite{hmst1}. Here we used
$k+k^\prime=(\sqrt{s},0)$,  ${\bf \hat e}=\bmk/|\bmk|$ and
$\hat{q}\equiv(\sqrt{s}-2m_t,0)$, $\sqrt{s}$ being the c.\,m.\,\,energy.
The result reads
\begin{equation}
\Delta\sigma_{\rm tot}^{\Gamma,1}=
2\,N_c\,\mbox{Im}\bigg\{
2i \big[\,C_V^{\rm born}\,C_V^{\rm bW,abs}+C_A^{\rm
  born}\,C_A^{\rm bW,abs}\,\big] G^0(a,v,m_t,\nu)
+\,\big[\,(C_V^{\rm born})^2+(C_A^{\rm born})^2\,\big]
\delta G^0_\Gamma(a,v,m_t,\nu)
\bigg\}
\,,
\label{dsignaNNLL}
\end{equation}
where $a\equiv -{\cal V}_c^{(s)}(\nu)/4\pi=C_F\alpha_s(m_t\nu)$. The term $G^0$ is
the zero-distance 
S-wave Green function of the non-relativistic Schr\"odinger
equation which is obtained from the LL terms in Eqs.\,(\ref{Lke}) and the
Coulomb potential. In dimensional regularization it has
the form~\cite{hmst1} 
\begin{eqnarray}
 G^0(a,v,m_t,\nu) =
 \frac{m_t^2}{4\pi}\left\{\,
 i\,v - a\left[\,\ln\left(\frac{-i\,v}{\nu}\right)
 -\frac{1}{2}+\ln 2+\gamma_E+\psi\left(1\!-\!\frac{i\,a}{2\,v}\right)\,\right]
 \,\right\}
 +\,\frac{m_t^2\,a}{4 \pi}\,\,\frac{1}{4\,\epsilon}
\,,
\label{deltaGCoul}
\end{eqnarray}  
where
$v  = 
\left((\sqrt{s}-2(m_t+\delta m_t)+i\Gamma_t)/m_t\right)^{1/2}$.
The term $\delta G^0_\Gamma$ represents the corrections originating
from the time dilatation correction in Eq.\,(\ref{Lke}) and reads
$\delta G^0_\Gamma(a,v,m_t,\nu)  = 
-i\frac{\Gamma_t}{2m_t}[
1+\frac{v}{2}\frac{\partial}{\partial v} + a\frac{\partial}{\partial a}
]G^0(a,v,m_t,\nu)$.
Note that the Wilson coefficients $C_{V/A}$ do not have a  LL anomalous
dimension, so only the matching conditions at $\nu=1$ appear in
Eq.\,(\ref{dsignaNNLL}). 

One can check that the terms proportional to
$C_{V/A}^{\rm bW,abs}$ in Eq.\,(\ref{dsignaNNLL}) are in agreement with the
full theory matrix elements from the interference
between the double-resonant amplitudes for the process
$e^+e^-\to t\bar t\to bW^+\bar b W^-$ (first diagram in
Fig.\,\ref{fig1}b) and the single-resonant amplitudes describing the
processes $e^+e^-\to t+\bar b W^-\to  bW^+\bar b W^- $  and 
$e^+e^-\to bW^+\bar t\to bW^+\bar b W^-$ in the
$t\bar t$ threshold limit for $m_t\to\infty$ (subsequent diagrams in
Fig.\,\ref{fig1}b). 
To find literal agreement between full and effective theory matrix elements
one has to replace the $i\epsilon$ terms in the resonant full theory top 
propagators by the Breit-Wigner term $i m_t\Gamma_t/2$.  As discussed
above, there are no further QCD corrections in the non-relativistic
limit due to the cancellation of the QCD interference effects caused by gluons
with ultrasoft momenta. 

The corrections given in Eq.\,(\ref{dsignaNNLL}) exhibit UV
$1/\epsilon$-divergences that have interesting features. Physically they arise from a
logarithmic high energy behavior of the top-antitop effective 
theory phase space integration for matrix elements containing a single
insertion of the Coulomb potential. Technically they enter the imaginary part
of the forward scattering amplitude because the imaginary parts of the
matching conditions of Eqs.\,(\ref{Cvaabs}) lead to a dependence on the real
part of $G^0$ (see Eq.~(\ref{deltaGCoul})). In the full theory this
logarithmic behavior is regularized by the top quark mass. It is important
that the divergences only exist because the top quark is not treated as
a stable particle. In particular, the UV divergences
from the time dilatation corrections arise from the Breit-Wigner-type high
energy behavior of the effective theory top propagator derived from
Eq.\,(\ref{Lke}), which differs from the one for a 
stable particle. Likewise, the interference effects described by the absorptive
electroweak matching conditions for the operators ${\cal O}_{V,\bbmp}$ and 
${\cal O}_{A,\bbmp}$ would not have to be accounted for if the top quarks
were stable particles. UV divergences of the same kind for the NNLL total
cross section have been observed and noted
before~\cite{hmst,hmst1,HoangTeubnerdist,HoangTeubner}, but no
resolution of the issue was provided.
From the point of view of having an EFT with non-hermitian contributions it is
obvious that these UV divergences must be handled in the canonical way using
renormalization. Since the divergences are directly related to operators with
non-hermitian Wilson coefficients, the renormalization procedure will naturally
involve operators with non-hermitian Wilson coefficients.  

The operators that are renormalized by the UV divergences displayed in
Eq.\,(\ref{dsignaNNLL}) are the two $(e^+e^-)(e^+e^-)$ operators 
\begin{eqnarray}
\tilde{\cal O}_V  = 
-\,\big[\,\bar e\,\gamma^\mu\,e\,\big]\,
\big[\,\bar e\,\gamma_\mu\,e\,\big]
\,,\qquad
\tilde{\cal O}_A  =
-\,\big[\,\bar e\,\gamma^\mu\,\gamma_5 \, e\,\big]\,
\big[\,\bar e\,\gamma_\mu\,\gamma_5\,e\,\big]
\,,
\end{eqnarray}
which give  the additional contribution 
$
\tilde\Delta {\cal L} = \tilde C_V \tilde{\cal O}_V 
+ \tilde C_A\tilde{\cal O}_A
$
to the effective theory Lagrangian,  $\tilde C_{V/A}$ being the Wilson
coefficients.
Because in this work we neglect QED effects, the electron and
positron act as classic fields and therefore  $\tilde C_{V}$ and $\tilde
C_{A}$ run only through mixing due to UV divergences such as in 
Eq.\,(\ref{dsignaNNLL}). 
Since only the imaginary parts of the coefficients $\tilde
C_{V/A}$ are relevant for the discussion,
we neglect the real contributions in the following.
Using the standard $\msb$ subtraction procedure to determine the (non-hermitian)
counterterms of the renormalized $\tilde{\cal O}_{V/A}$ operators and
standard methods to compute and to solve the anomalous dimensions, one
obtains the following form of the Wilson coefficients 
$\tilde C_{V/A}$ for scales below $m_t$ ($v<1$),
\begin{eqnarray}
\tilde C_{V/A}(\nu)& = & \tilde C_{V/A}(1) + 
i\,\frac{2 N_c m_t^2 C_F}{3\beta_0}\,\bigg\{
\bigg[ \Big( (C_{V/A}^{\rm born})^2+ (C_{V/A}^{\rm ax})^2 \Big)\,\frac{\Gamma_t}{m_t}
 + 3 C_{V/A}^{\rm born} C_{V/A}^{\rm bW,abs}\bigg]\,\ln(z)
\nonumber\\[2mm] & &\qquad
 - \frac{4C_F}{\beta_0}\,\frac{\Gamma_t}{m_t}\,(C_{V/A}^{\rm born})^2\ln^2(z)
 + \frac{4(C_A+2C_F)}{\beta_0}\,\frac{\Gamma_t}{m_t} \,(C_{V/A}^{\rm born})^2\rho(z)
\bigg\}
\,,
\label{tildeCav}
\end{eqnarray}
where $z  \equiv  \alpha_s(m_t\nu)/\alpha_s(m_t)$, 
$\rho(z)  =  \frac{\pi^2}{12}-\frac{1}{2}\ln^22 + \ln2 \ln(z) - {\rm
  Li}_2\left(\frac{z}{2}\right)$
and the $\tilde C_{V/A}(1)$ are the hard matching conditions.
The operators $\tilde {\cal O}_{V/A}$ lead to an additional
contribution to the total cross section of the form
\begin{eqnarray}
\Delta\sigma_{\rm tot}^{\Gamma,2} & = & 
\mbox{Im}\Big[\,{\tilde C_V} +{\tilde C_A}\,\Big]
\,.
\end{eqnarray}
For details of the computations see Ref.~\cite{HoangReisser}.
Parametrically $\Delta\sigma_{\rm tot}^{\Gamma,2}$ is of order
$g^6$ 
and thus constitutes a NLL contribution as one can also see from
the fact that the corresponding UV divergences were generated in NNLL order
effective theory matrix
elements. The correction $\Delta\sigma_{\rm tot}^{\Gamma,2}$ is
energy-independent, but it is scale-dependent and
compensates the logarithmic scale-dependence in the NNLL contribution
$\Delta\sigma_{\rm tot}^{\Gamma,1}$. 
The matching conditions $\tilde C_{V/A}(\nu=1)$ are presently
unknown and will be analysed in subsequent work. For now we set them to zero
in the numerical analysis presented below. We note that it was shown
in~\cite{HoangTeubnerdist} that the difference between the full theory phase
space (which is cut off by the large, but finite $m_t$) and the effective theory
phase space (which is infinite in the computation of the forward scattering
amplitude) contributes to $\tilde C_{V/A}(\nu=1)$ and also represents a NLL
effect.

\section{Numerical Analysis}
\label{sectionanalysis}
%
%
%
%
\begin{figure}[t] 
 \leavevmode
 \epsfxsize=10cm
 \leavevmode
 \epsffile[100 430 580 730]{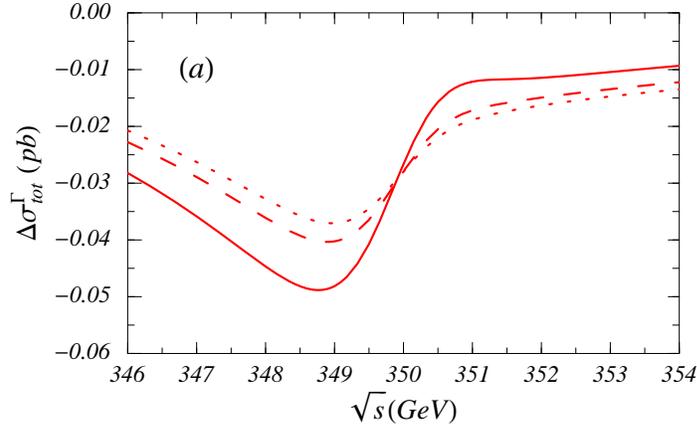}
%
 \caption{
The sum $\Delta\sigma_{\rm tot}^{\Gamma,1}+\Delta\sigma_{\rm tot}^{\Gamma,2}$
in $pb$ for $M_{\rm 1S}=175$~GeV, 
$\alpha=1/125.7$, $s_w^2=0.23120$, $V_{tb}=1 $, $M_W=80.425$~GeV,
$\Gamma_t=1.43$~GeV and $\nu=0.1$ (solid
curves),  $0.2$ (dashed curves) and $0.3$ (dotted curves) in the energy
range $346~\mbox{GeV} < \sqrt{s} < 354$~GeV.
 \label{figanalysis} }
\end{figure}

In Fig.\,\ref{figanalysis} we have plotted 
the sum of
$\Delta\sigma_{\rm tot}^{\Gamma,1}$  and $\Delta\sigma_{\rm tot}^{\Gamma,2}$ 
in picobarn in the 1S mass scheme~\cite{HoangTeubnerdist} for
$M_{\rm 1S}=175$~GeV, the fine structure constant $\alpha=1/125.7$, $s_w^2=0.23120$, $V_{tb}=1$ and
$M_W=80.425$~GeV with the renormalization scaling parameter $\nu=0.1$ (solid
curves),  $0.2$ (dashed curves) and $0.3$ (dotted curves). 
For the QCD
coupling we used $\alpha_s(M_Z)=0.118$ as an input and employed 4-loop
renormalization group running. Note that in the 1S scheme 
$\delta m_t=M_{\rm 1S}({\cal V}_c^{(s)}(\nu)/4\pi)^2/8$. 
For the top 
quark width we adopted the value $\Gamma_t=1.43$~GeV. Note that
in a  complete analysis of electroweak effects the top quark
width depends on the input parameters given above and is not an independent
parameter. For the purpose of the numerical analysis in this work, however,
our treatment is sufficient.
We find that the sum of the corrections is negative and shows a moderate $\nu$
dependence. We find that the corrections are around $-10\%$
for energies below the peak, between $-2\%$ and $-4\%$ close to the peak and
about $-2\%$ above the peak. Interestingly, they partly compensate  
the sizeable positive QCD corrections found
in~\cite{Hoang3loop,HoangEpiphany}. The peculiar energy dependence of the
corrections, caused by the dependence on the real part of the Green function
$G^0$, also leads to a slight displacement of the peak position. Relative to
the peak position of the LL cross section one obtains a shift of $(30,35,47)$~MeV
for $\nu=(0.1,0.2,0.3)$. This shift is comparable to the expected
experimental uncertainties of the top mass measurements from the
threshold scan~\cite{TTbarsim}.


\end{document}